\documentclass[a4paper,fleqn,usenatbib]{mnras}
\usepackage{graphicx}	
\usepackage{amsmath}	
\usepackage{amssymb}	
\usepackage{multicol}        
\usepackage{bm}		
\usepackage{pdflscape}	
\usepackage{color}
\usepackage{multirow}
\usepackage{textcomp}
\usepackage{amssymb}

\usepackage[T1]{fontenc}
\usepackage{ae,aecompl}

\usepackage[utf8]{inputenc}

\def\be{\begin{equation}}
\def\ee{\end{equation}}

\title[Magnetoionic environments of FRBs]{On the magnetoionic environments of fast radio bursts}

\author[Wei-Yang Wang et al.]{
Wei-Yang Wang$^{1,2,3}$,
Bing Zhang$^{4}$\thanks{E-mail: zhang@physics.unlv.edu}, Xuelei Chen$^{1,2,5}$\thanks{E-mail: xuelei@cosmology.bao.ac.cn}, Renxin Xu$^{3,6}$\\
$^1$Key Laboratory of Computational Astrophysics, National Astronomical Observatories, Chinese Academy of Sciences, Beijing 100012, China\\
$^2$University of Chinese Academy of Sciences, Beijing 100049, China\\
$^3$School of Physics and State Key Laboratory of Nuclear Physics and Technology, Peking University, Beijing 100871, China\\
$^4$Department of Physics and Astronomy, University of Nevada, Las Vegas, NV 89154, USA\\
$^5$Center for High Energy Physics, Peking University, Beijing 100871, China\\
$^6$Kavli Institute for Astronomy and Astrophysics, Peking University, Beijing 100871, China}

\date{Accepted XXX. Received YYY; in original form ZZZ}

\pubyear{2020}

\begin{document}
\label{firstpage}
\pagerange{\pageref{firstpage}--\pageref{lastpage}}
\maketitle

\begin{abstract}
Observations of the Faraday rotation measure, combined with the dispersion measure, can be used to infer the magnetoionic environment of a radio source.
We investigate the magnetoionic environments of FRBs by deriving their estimated average magnetic field strengths along the line of sight $\langle B_{\parallel}\rangle$ in their host galaxies and comparing them with those of Galactic pulsars and magnetars. We find that for those FRBs with RM measurements, the mean $\langle B_{\parallel}\rangle$ is $1.77^{+9.01}_{-1.48}\,\rm\mu G$ and $1.74^{+14.82}_{-1.55}\,\rm\mu G$ using two different methods, which is slightly larger but not inconsistent with the distribution of Galactic pulsars, $1.00^{+1.51}_{-0.60}\,\rm\mu G$. Only six Galactic magnetars have estimated $\langle B_{\parallel}\rangle$. Excluding PSR J1745--2900 that has an anomalously high value due to its proximity with the Galactic Centre, the other three sources have a mean value of $1.70\,\rm\mu G$, which is statistically consistent with the  $\langle B_{\parallel}\rangle$ distributions of both Galactic pulsars and FRBs. There is no apparent trend of evolution of magnetar $\langle B_{\parallel}\rangle$ as a function of age or surface magnetic field strength. Galactic pulsars and magnetars close to the Galactic Centre have relatively larger $\langle B_{\parallel}\rangle$ values than other pulsars/magnetars. We discuss the implications of these results for the magnetoionic environments of FRB 121102 within the context of magnetar model and the model invoking a supermassive black hole, and for the origin of FRBs in general. 

\end{abstract}

\begin{keywords}
pulsars: general - stars: neutron - radio continuum: transients
\end{keywords}

\section{Introduction}

Fast radio bursts (FRBs) are millisecond-duration coherent emissions of extragalactic/cosmological origin \citep[e.g.][]{Lorimer07,Thornton13,Chatterjee17,Bannister19,Prochaska19,Marcote20}. The physical origin(s) of these events are unknown.

The polarization properties of FRBs may shed light on the magneto-ionic environment and the physical mechanisms of FRBs.
Only a small fraction of FRBs have polarization measurements, but the data show a perplexing picture \citep{Petroff19}: whereas some bursts show a nearly 100\% linear polarization percentage, some others have moderate or even negligible polarization degrees. Faraday rotation measures (RMs) can be measured for those events with linear polarizations. 
The first observed repeating source, FRB 121102, 
has a very large, varying Faraday rotation measure (RM) of the order of $10^5 \ {\rm rad \ m^{-2}}$ \citep{Chatterjee17,Marcote17,Michilli18}. On the other hand, other sources, either apparently non-repeating (FRB 180924, FRB 181112 and FRB 190102, \citealt{Bannister19,Prochaska19,Macquart20}) or repeating (FRB 180301 and FRB 180916.J0158+65, \citealt{theCHIME19,Luo20}), show a much smaller (by 2-4 orders of magnitude) RM.

When combined with the dispersion measure (DM), the RM can be used to infer the average magnetic field strength along the line of sight (LOS) in the host galaxy of the FRB. For FRB 121102, this is $B_{\parallel}\sim1$ mG, which is several orders of magnitude higher than that of the interstellar medium \citep{Michilli18}. In the literature, two scenarios have been discussed to account for such an extreme magnetoionic environment. One scenario invokes a strongly magnetized neutron star (or magnetar), which injects a highly magnetized wind to the medium forming a magnetar wind nebula (MWN). The large RM of FRB 121102 may be interpreted within the framework of such a MWN \citep{Piro18,Metzger19}. The second scenario invokes a source in the vicinity of a supermassive black hole \citep{Michilli18,Zhang18a}, the only known location where an extremely high RM has been observed \citep{Eat13}. 

Motivated by the recent intriguing observations of an FRB-like burst associated with a hard X-ray burst from a Galactic magnetar SGR 1935+2154 \citep{STARE2,CHIME20,HXMT20,integral20,konus20,agile20}, we compare the $B_{\parallel}$ values derived from FRB observations with those of Galactic magnetars and pulsars as an effort to diagnose the origin of $B_{\parallel}$ of FRBs. The results are presented in Section 2. The implications of the results on the origin of FRBs are discussed in Section 3. The results are summarized in Section 4.

\section{Magnetoionic environments of Galactic pulsars, magnetars, and FRBs}

\subsection{Galactic pulsars and magnetars}

\begin{figure*}
\begin{center}
\includegraphics[width=0.98\textwidth]{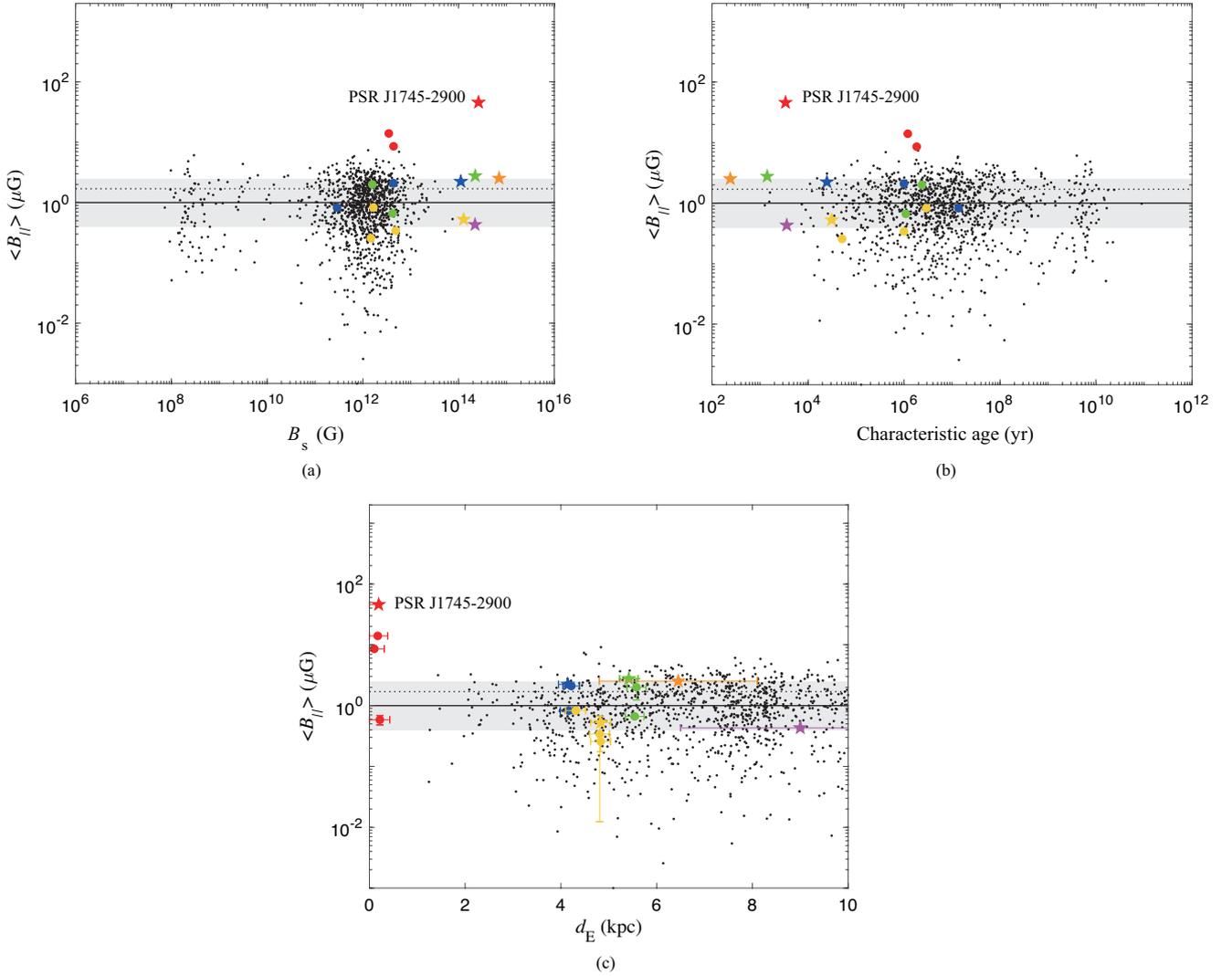}
\caption{\small{(a) Top left: The parallel magnetic field strength $\langle B_{\parallel}\rangle$ of Galactic pulsars (dots) and magnetars (pentagrams) as a function of their surface magnetic field strength $B_{\rm s}$. The six magnetars are denoted by different colors. The dots with the same colors are normal pulsars that are within $0.5$ kpc from the magnetar for the corresponding color. The data are quoted from \citet{Manchester05}. The dashed and dotted lines are the mean values of $\langle B_{\parallel}\rangle$ for pulsars and magnetars (except the Galactic central magnetar), respectively. The grey zone is the 1-$\sigma$ region of the pulsar $\langle B_{\parallel}\rangle$ distribution;
(b) Top right: Same as the top left panel, but plotted as a function of pulsar age;
(c) Bottom: Same as the top left but plotted as a function of the distance from the Galactic Centre. The galactic centre distance from Earth is adopted as $d_{\rm gc}=8.32$ kpc \citep{Gillessen17};
}}
\label{fig1}
\end{center}
\end{figure*}

\begin{figure*}
\begin{center}
\includegraphics[width=0.9\textwidth]{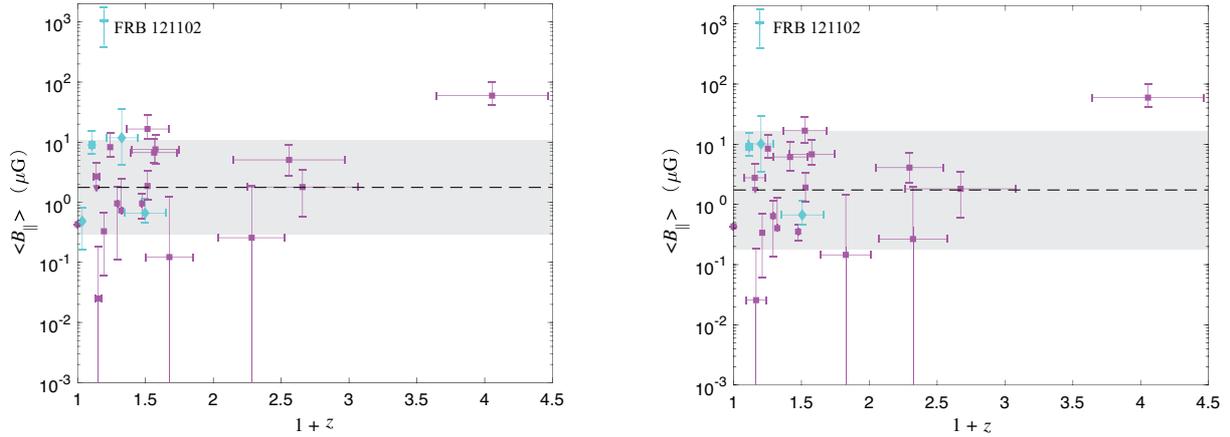}
\caption{\small{The parallel magnetic field strength $\langle B_{\parallel}\rangle$ of FRBs within their host galaxies. Left: Subtracted the Galactic DM contribution by the method of NE2001. The light blue crosshairs in the upper-left corner is for FRB 121102. 
The light blue diamonds are other repeating FRBs. The purple squares are apparently non-repeating FRBs. The black dashed line is the mean value of $\langle B_{\parallel}\rangle$ for FRBs except FRB 121102. The grey zone is the 1-$\sigma$ region of the distribution. Right: Same as the left but for the method of YMW16.}}
\label{fig2}
\end{center}
\end{figure*}

\begin{figure*}
\begin{center}
\includegraphics[width=0.9\textwidth]{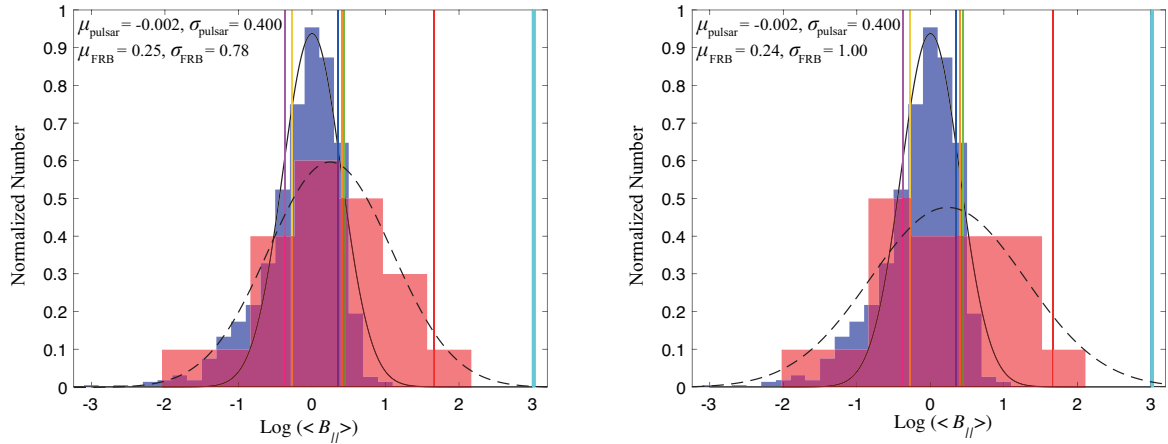}
\caption{\small{Histograms of the log normal distribution of $\langle B_{\parallel}\rangle$ for FRBs (red) and pulsars (blue). The colored solid lines are magnetars: SGR 1935+2154 (purple), XTE J1810--197 (yellow), PSR J1622--4950 (blue), Swift J1818.0--1607 (orange), 1E 1547.0--5408 (green) and PSR J1745--2900 (red). The black solid and dashed lines are the best fitting for FRBs and pulsars. The light blue vertical line shows the $\langle B_{\parallel}\rangle$ of FRB 121102. Left: The Galactic DM contribution is subtracted using NE2001; Right: The Galactic DM contribution is subtracted using YMW16.}}
\label{fig3}
\end{center}
\end{figure*}

The average magnetic field strength (i.e. the absolute value) along the line of sight (LOS) may be derived by combining the measurements of RM and DM of radio pulses \citep[e.g.][]{Manchester72,Manchester74,Han06,Nou08,Han18}):
\begin{equation}
\langle B_{\parallel}\rangle =1.23\left|{\rm\frac{RM}{DM}}\right|\,\rm{\mu G},
\label{eq0}
\end{equation}
where RM is in units of $\rm rad\,m^{-2}$ and DM is the dispersion measure in units of $\rm pc\,cm^{-3}$.
For Galactic pulsars (data are quoted from \citealt{Manchester05}), the contribution to RM from the plasma associated with the source is negligible, so that $\langle B_{\parallel} \rangle$ can be used to map the magnetic field structure of the Galaxy \citep{Hanqiao94}.
The RM term $\rm RM_{iono}$ originated from the Earth's ionosphere \citep{S-B13} is $0.5-3\rm\,rad\,m^{-2}$. The
RMs for most Galactic pulsars are larger than $\rm 10\,rad\,m^{-2}$.
For pulsars that have $\rm DM \sim 100 pc cm^{-3}$, the variability of the ionospheric RM only slightly affects $\langle B_{\parallel}\rangle$ and can, hence be neglected.
The $\langle B_{\parallel}\rangle$ of Galactic pulsars are calculated with Equation (\ref{eq0}) and plotted in Figure \ref{fig1}.

We consider whether Galactic magnetars can host a more extreme magneto-ionic environment than pulsars. According to theory, a young magnetar may power an MWN, which may be more magnetized than a regular pulsar wind nebula. 
In an expanding MWN, the DM and RM would generally decrease with time \citep{Metzger17,Yang17,Piro18,Margalit18,Metzger19}, suggesting that one may observe a trend of decreasing $\langle B_{\parallel}\rangle$ with time or surface magnetic field strength.
Among over twenty observed magnetars, there are six sources that emitted polarized radio waves to allow $\langle B_{\parallel}\rangle$ to be measured (\citealt{CHIME20,Lower20}, also see \citealt{Kaspi17} for a review).

In Figure \ref{fig1}(a) and (b), we plot $\langle B_{\parallel}\rangle$ as a function of magnetar surface magnetic field $B_{\rm s}$ and characteristic age for these six radio magnetars. 
The characteristic age is close to real age under the assumptions that the initial spin period of the pulsar is much less than its current period and that pulsar spindown is dominated by dipolar magnetic braking. The surface magnetic field configuration is likely more complicated but the exact strength of the surface field is difficult to infer from observations. For simplicity, we use $B_{\rm S}$ as a proxy of the magnetar surface magnetic field and study how $\langle B_{\parallel}\rangle$  depends on it.
For comparison, we also select some pulsars that are within 0.5~kpc from the magnetars. The distance between a pulsar and a magnetar is calculated by combining their radial distances inferred from DM measurements and their transverse distance measured based on their celestial coordinates. 
The distance to Swift J1818.0--1607 is estimated to be 4.8 kpc (according to the YMW16 electron density model, \citealt{YMW16}) or $8.1\pm1.6$ kpc (according to NE2001 electron density model, \citealt{NE2001}).
Another source, SGR 1935+2154, has an unknown distance, but one may assume that the SGR is physically related to a supernova remanent SNR G57.2+0.8, which has a distance of $9.0\pm2.5$ kpc \citep{Zhong20,Zhou20}.

Figure \ref{fig1}(a) shows that except the magnetar PSR J1745--2900, whose abnormally large $\langle B_{\parallel}\rangle$ may be related to its proximity with the Galactic super-massive black hole, the other five magnetars have $\langle B_{\parallel}\rangle$ consistent with the pulsar $\langle B_{\parallel}\rangle$ distribution (two within $1\sigma$, one slightly outside the $1\sigma$ region). Indeed some regular pulsars have larger $\langle B_{\parallel}\rangle$ values than magnetars. Compared with normal pulsars within 5 kpc, the five magnetars also do not have systematically higher $\langle B_{\parallel}\rangle$ values. Figure \ref{fig1}(b) plots $\langle B_{\parallel}\rangle$ against pulsar age. One can see that PSR J1745--2900 is not the youngest but has the highest $\langle B_{\parallel}\rangle$, further suggesting that its abnormal $\langle B_{\parallel}\rangle$ is related to its special environment. For the five magnetars, there is also no trend of decreasing $\langle B_{\parallel}\rangle$ with characteristic age.

In view of the abnormally high $\langle B_{\parallel}\rangle$ of the magnetar PSR J1745--2900, we also investigate the relationship between $\langle B_{\parallel}\rangle$ 
and the distance of pulsars from the Galactic Centre. 
The results are shown in Figure \ref{fig1}(c).
The $\langle B_{\parallel}\rangle$ values for the magnetar PSR J1745--2900 and two pulsars close to it (PSR J1746--2849 and PSR J1746--2856) have $\langle B_{\parallel}\rangle$ significantly exceeding those of Galactic pulsars, further suggesting that it is the special location near the Galactic centre that caused the abnormally high $\langle B_{\parallel}\rangle$.
The Galactic Centre super-massive black hole likely provides the strong local magnetic field.
The magnetar PSR J1745--2900 indeed shows a larger $\langle B_{\parallel}\rangle$ than the pulsars. This may suggest that the magnetar may contribute to an additional $\langle B_{\parallel}\rangle$ value. However, due to the large uncertainty of the distance measurements from the Galactic Centre (noticing the large error bars of the red dots), this conclusion cannot be firmly drawn. It is interesting to note that at distances beyond 1 kpc, $\langle B_{\parallel}\rangle$ is essentially independent of distance from the Galactic Centre.

The $\langle B_{\parallel}\rangle$ distribution of Galactic pulsars can be fitted with a log normal function, as shown in Figure \ref{fig3}.
The mean value of log$\langle B_{\parallel}\rangle$ is $\mu_{\rm pulsar}=-0.002$ with a standard deviation of $\sigma_{\rm pulsar}=0.40$ for the log normal distribution.
Excluding the Galactic Centre magnetar PSR J1745--2900, the other five magnetar sources have a mean $\langle B_{\parallel}\rangle$ of $1.84\,\rm\mu G$, which is 
statistically consistent with the pulsar $\langle B_{\parallel}\rangle$ distribution.

\subsection{FRBs}
For a source at cosmological distances, the observed total $\rm RM_{obs}$ consists of contributions from several different terms,
\begin{equation}
\rm RM_{obs}=RM_{iono}+RM_{Gal}+RM_{IGM}+RM_{HG, sr},
\end{equation}
where $\rm RM_{Gal}$ is the Galactic component, $\rm RM_{IGM}$ is contributed from the intergalactic medium (IGM), and $\rm RM_{HG, sr}$ is contributed by the host galaxy and the FRB source as measured in the Observer's frame. The true value of the latter in the rest frame of the host galaxy is
\begin{equation}
{\rm RM^{Loc}_{HG, sr}}={\rm RM_{HG, sr}}{(1+ z)^2},
\label{eq1}
\end{equation}
where $z$ is the redshift. 

In order to determine $\rm RM_{Gal}$, we identify the RM of the known NRAO VLA Sky Survey (NVSS) sources \citep{Taylor09}.
The closest sources are $\lesssim 2 ^{\circ}$ away from the position of FRBs. For the FRB sources located in the survey blind regions, we identify the RM based on their closest pulsars \citep{Han18}.
Alternatively, a simulation result of Galactic RM is given by \cite{Oppermann15}, even though the simulation results are based on some inputs and assumptions.
The strength of intergalactic magnetic fields is much lower. 
A safe upper limit is
$\sim10^{-9}$\,G \citep{Dai02,Ando10,Dermer11,Arlen14}, which gives $\rm RM_{IGM} \lesssim 1\,\rm rad\,m^{-2}$, so that it can usually be neglected.
After subtracting the contributions from $\rm RM_{iono}$ and $\rm RM_{Gal}$, one can finally derive $\rm RM_{HG, sr}$ and $\rm RM^{Loc}_{HG, sr}$.

The observed DM also consists of several terms:
\begin{equation}
\rm DM_{obs}=DM_{Gal}+DM_{halo}+DM_{IGM}+DM_{HG, sr},
\label{eq:DM}
\end{equation}
where $\rm DM_{Gal}$ can be derived from the Galactic electron density models.
However, the two well-known Galactic electron density models NE2001 and YWM16, give quite different results sometimes \citep{NE2001,YMW16}.
In the following discussion, we consider the electron density models, separately.
The DM associated with the Milky Way halo is adopted as $\rm DM_{halo}=30\pm15\,pc\,cm^{-3}$ according to \cite{Dolag15}.
The average value of the IGM component is \citep{Deng14,Zhang18b}
\begin{equation}
{\rm DM}_{\rm IGM} = \frac{3cH_0\Omega_bf_{\rm IGM}}{8\pi G m_p}\int^z_0\frac{\chi(z)(1+z)dz}{[\Omega_m(1+z)^3+\Omega_{\lambda}]^{\frac{1}{2}}},
\label{eq:IGM}
\end{equation}
where the free electron number per baryon in the universe is $\chi(z) \approx 7/8$ and the fraction of baryons $f_{\rm IGM}\sim0.83$.
For a distant cosmological FRB, the DM is mainly contributed by IGM rather than Milky Way or the host galaxy, in contrast to the RM that has a significant contribution from the host.
The $\Lambda$CDM cosmological parameters are taken as $\Omega_m =0.315 \pm 0.007, \Omega_b h^2 = 0.02237 \pm 0.00015$, and $H_0=67.36\rm \pm 0.54\,\rm km\,s^{-1}Mpc^{-1}$ \citep{Planck18}. 
The remaining term in Eq.(\ref{eq:DM}) is $\rm DM_{HG, sr}$ in the observer frame. The intrinsic value in the source frame is related to it by
\begin{equation}
{\rm DM^{Loc}_{HG, sr}} = (1+z){\rm DM_{HG, sr}}.\label{eq:host}
\end{equation}

In practice, $\rm DM_{HG, sr}$ is difficult to derive from the observed $\rm DM_{obs}$, even if the redshift of the FRB is precisely known because there is a large scatter of $\rm DM_{IGM}$ around Eq.(\ref{eq:IGM}) due to the large scale structure fluctuation \citep{McQuinn14}.
Several FRB sources have measured redshifts, so that their $\rm DM_{IGM}$ and uncertainties can be derived \citep{McQuinn14}. As a result, we can calculate $\rm DM_{HG, sr}$ and its uncertainty $\rm \delta DM_{HG, sr}$ directly.
For the sources without precise redshift measurements, we adopt the opposite approach, to consider the distribution of $\rm DM^{Loc}_{HG, sr}$ based on host galaxy models. 
Following a generic constraint by \cite{Li20}, we assume that $\rm DM^{Loc}_{HG, sr}=85\pm35\rm\,pc\,cm^{-3}$, which is consistent with the results of
the average DM contribution from the host galaxy in the local frame e.g., \cite{Xu15} and \cite{Luo18}.

The average magnetic field for the host galaxy along the LOS can then be derived as 
\begin{equation}
\langle B_{\parallel}\rangle
=1.23(1+z)\left|{\rm\frac{RM_{HG, sr}}{DM_{HG, sr}}}\right|\,\rm{\mu G},
\label{eq3}
\end{equation}
where for the redshift we either adopted the spectroscopic value if measured, or estimated using Eq.(\ref{eq:host}) with error properly introduced. 

Based on the FRB catalog (http://frbcat.org/, \citealt{Petroff16}), in Table \ref{tab1} we list all the FRBs with both RM and DM measured.
According to Eq.(\ref{eq3}), we estimate $\langle B_{\parallel}\rangle$ for these FRBs. 
The results are listed in Tables 1 and 2, where $\rm DM_{Gal}$ is derived using the NE2001 and YMW16 models, respectively.
Several sources have $\rm DM_{HG, sr}$ smaller than $\rm DM_{IGM}$ and $\rm\delta DM_{IGM}$, which differ from FRB 121102 whose $\rm DM_{HG, sr}$ and $\rm DM_{IGM}$ are comparable.
We show the lower limits for these sources.
The distribution of $\langle B_{\parallel}\rangle$ is also presented in Figure \ref{fig3}, which can be also fitted with a log normal function when FRB 121102 is excluded.
The mean value derived by the two Galactic electron density models are $\mu_{\rm FRB}=0.25,\,\sigma_{\rm FRB}=0.78$ (NE2001) and $\mu_{\rm FRB}=0.24,\,\sigma_{\rm FRB}=1.00$ (YMW16), respectively. These values are slightly higher but not inconsistent with the distributions of both Galactic pulsars and magnetars.

\begin{table*}
\begin{center}
\caption{DM (subtracted by NE2001) and RM of FRB and magnetar sample}
\begin{tabular}{ccccccc}
\hline \hline
Source$^a$ & $\rm DM_{excess}^b$ & $z^{c}$ & $\rm DM^d_{HG,sr}$ & $\rm RM_{obs}$ & $\rm RM_{HG,sr}$ & $\langle B_{\parallel}\rangle$\\
 & $\rm(pc\,cm^{-3})$ & & $\rm(pc\,cm^{-3})$ & $\rm(rad\,m^{-2})$ & $\rm(rad\,m^{-2})$ & $(\rm\mu G)$ \\
\hline
FRB 121102 & 374 & $0.193$ & $140\pm85$ & $(0.9-1.0)\times10^5$ & $(0.9-1.0)\times10^5$ & $989.4-1084.1$\\
FRB 180916.J0158+65 & 119 & $0.034$ & $73.8\pm15$ & $-114\pm0.6$ & $200.6\pm17.7$ & $0.5^{+0.3}_{-0.3}$\\
FRB 180924 & 290.92 & $0.321$ & $15.8$ & $14\pm1$ & $7.1\pm15.4$ & $0.7_{-0.7}^{+1.7}$\\
FRB 181112 & 457.27 & $0.476$ & $42.9$ & $10.9\pm0.9$ & $-22.5\pm5.8$ & $1.0^{+0.4}_{-0.4}$\\
FRB 190102 & 276.3 & $0.291$ & $28.3$ & $110$ & $-17\pm12$ & $1.0^{+0.8}_{-0.8}$\\
\hline
FRB 110523 & 549.78 & $0.56\pm0.17$ & $54.3^{+31.8}_{-25.5}$ & $-186.1\pm1.4$ & $-188.6\pm19.3$ & $6.7^{+4.8}_{-2.2}$\\
FRB 150215 & 648.4 & $0.68\pm0.18$ & $50.7^{+29.2}_{-23.7}$ & $-3.3\pm12.2$ & $3.0\pm27.3$ & $0.1^{+1.1}_{-0.1}	$ \\
FRB 150418 & 557.7 & $0.57\pm0.17$ & $54.0^{+31.7}_{-25.4}$ & $36\pm52$ & $-211\pm59$ & $7.6^{+5.8}_{-3.2}$\\
FRB 150807 & 199.6 & $0.15\pm0.02$ & $73.8^{+32.5}_{-31.3}$ & $12.0\pm7$ & $-1.3\pm8$ & $0.03_{-0.03}^{+0.15}$\\
FRB 160102 & 2553.1 & $3.05\pm0.41$ & $21.0^{+12.0}_{-9.8}$ & $-220.6\pm6.4$ &$-249.3\pm12.7$ & $59.5^{+42.2}_{-18.6}$\\
FRB 171209 & 1414.4 & $1.56\pm0.41$ & $33.2^{+22.6}_{-16.4}$ & $121.6 \pm4.2$ & $115.5\pm9.2$ & $5.1^{+4.0}_{-2.4}$\\
FRB 180301 & 342 & $0.33\pm0.11$ & $64.2^{+35.0}_{-29.4}$ & $520-570$ & $504.5-546.2$ & $9.7-13.9$\\
FRB 180309 & 188.73 & $0.14\pm0.02$ & $74.8^{+32.8}_{-31.6}$ & $<150$ & $142.8<$ & $<2.7^{+1.9}$\\
FRB 180311 & 1495.7 & $1.66\pm0.41$ & $32.0^{+21.4}_{-15.7}$ & $4.8\pm7.3$ & $-17.5\pm10.3$ & $1.8^{+1.7}_{	-1.2}$\\
FRB 180714 & 1180.92 & $1.28\pm0.25$ & $37.2^{+21.7}_{-17.5}$ & $-25.9\pm5.9$ & $55.2\pm21.4$ & $0.3^{+1.6}_{-0.3}$\\
FRB 190303.J1353+48 & 163.4 & $0.11\pm0.02$ & $76.9^{+33.8}_{-32.5}$ & $-504.4\pm0.4$ & $498.5\pm12$ & $9.0^{+6.3}_{-2.6}$\\
FRB 190604.J1435+53 & 490.65 & $0.50\pm0.15$ & $56.8^{+32.3}_{-26.4}$ & $-16\pm1$ & $15.7\pm1.4$ & $0.7^{+0.5}_{-0.2}$\\
FRB 190608 & 271. 5 & $0.24\pm0.10$ & $68.5^{+36.7}_{-31.2}$ & $353\pm2$ & $370.3\pm9.6$ & $8.3^{+5.8}_{-2.5}$\\
FRB 190611 & 233.57 & $0.19\pm0.09$ & $71.2^{+38.5}_{-32.6}$ & $20\pm4$ & $15.9\pm12.1$ & $0.3^{+0.3}_{-0.3}$\\
FRB 190711 & 506.7	 & $0.52\pm0.16$ & $56.1^{+32.2}_{-26.2}$ & $9\pm2$ & $56.05\pm16$ & $1.9^{+1.4}_{-0.8}$\\
FRB 191108 & 506.1 & $0.51\pm0.16$ & $56.1^{+32.2}_{-26.2}$ & $474\pm3$ & $497.7\pm22.6$ & $16.6^{+11.8}_{-5.2}$\\
\hline
1E 1547.0--5408 & -  & - & 830 & $-1860\pm4$ & - & $2.8^{+0.2}_{-0.2}$\\
PSR J1622--4950 & -  & - & 820 & $-1484\pm1$ & - & $2.2^{+0.2}_{-0.2}$\\
PSR J1745--2900 & -  & - & 1778 & $-66080\pm24$ & - & $45.9^{+0.1}_{-0.1}$\\
XTE J1810--197 & -  & - & 178 & $76\pm1$ & - & $0.5^{+0.01}_{-0.01}$\\
Swift J1818.0--1607 & -  & - & 706 & $1442\pm0.2$ & - & $2.5^{+0.0003}_{-0.0003}$\\
SGR 1935+2154/FRB 200428 & -  & - & 333.7 & $116\pm7$ & - & $0.4^{0.03}_{-0.03}$\\
\hline \hline
\end{tabular}
\label{tab1}
\end{center}
$^{\rm a}${FRB 121102, FRB 180301, FRB 180916.J0158+65, FRB 190303.J1353+48 and FRB 190604.J1435+53 are repeaters.\\}
$^{\rm b}$$\rm DM_{excess}$ is identified as $\rm DM_{IGM}+DM_{HG, sr}$.\\
$^{\rm c}${FRB 121102, FRB 180916.J0158+65, FRB 180924 and FRB181112 have been localized in their host galaxies. Their redshifts are measured from their host galaxies. For other sources, their redshifts are calculated from Equation (\ref{eq:host}), with the assumption of $\rm DM^{Loc}_{HG, sr}=85\pm35\,pc\,cm^{-3}$ \citep{Li20}. The redshift errors are derived by the $\delta \rm DM_{IGM}$ from the IGM fluctuation \citep{McQuinn14}. For FRB 160102, FRB 180311 and FRB 171209, we assume $\rm \delta DM_{IGM}=350\,pc\,cm^{-3}$ due to their redshifts are larger than 1.4.\\}
$^{\rm d}${The uncertainty of $\rm DM_{HG, sr}$ for FRB 121102 is $85\rm\,pc\,cm^{-3}$ \citep{Ten17}. FRB 180916.J0158+65 is too close so that we let the $\rm\delta DM_{HG, sr}=15\rm\,pc\,cm^{-3}$. For other localized bursts, the uncertainty is larger than the mean value.\\}
References: {\cite{Bannister19}, \cite{Caleb18}, \cite{CHIME20}, \cite{Chatterjee17}, \cite{theCHIME19}, \cite{Connor20}, \cite{Day20}, \cite{Fonseca20}, \cite{Keane16}, \cite{Luo20}, \cite{Macquart20}, \cite{Manchester05}, \cite{Marcote17}, \cite{Marcote20}, \cite{Masui15}, \cite{Michilli18}, \cite{Osl19}, \cite{Petroff17}, \cite{Prochaska19}, \cite{Ravi16}, \cite{Ten17}}
\end{table*}

\begin{table*}
\begin{center}
\caption{DM (subtracted by YMW16) and RM of FRB and magnetar sample}
\begin{tabular}{ccccccc}
\hline \hline
Source & $\rm DM_{excess}$ & $z$ & $\rm DM_{HG,sr}$ & $\rm RM_{obs}$ & $\rm RM_{HG,sr}$ & $\langle B_{\parallel}\rangle$\\
 & $\rm(pc\,cm^{-3})$ & & $\rm(pc\,cm^{-3})$ & $\rm(rad\,m^{-2})$ & $\rm(rad\,m^{-2})$ & $(\rm\mu G)$ \\
\hline
FRB 121102 & 374 & $0.193$ & $140\pm85$ & $(0.9-1.0)\times10^5$ & $(0.9-1.0)\times10^5$ & $989.4-1084.1$\\
FRB 180916.J0158+65$^a$ & 19.2 & $0.034$ & $-$ & $-114\pm0.6$ & $200.6\pm17.7$ & $-$\\
FRB 180924 & 303.77 & $0.321$ & $28.7$ & $14\pm1$ & $7.1\pm15.4$ & $0.7_{-0.7}^{+1.7}$\\
FRB 181112 & 530.24 & $0.476$ & $115.8$ & $10.9\pm0.9$ & $-22.5\pm5.8$ & $1.0^{+0.4}_{-0.4}$\\
FRB 190102 & 290.29 & $0.291$ & $42.3$ & $110$ & $-17\pm12$ & $1.0^{+0.8}_{-0.8}$\\
\hline
FRB 110523 & 560.3 & $0.58\pm0.17$ & $53.9^{+31.2}_{-25.2}$ & $-186.1\pm1.4$ & $-188.6\pm19.3$ & $6.8^{+4.9}_{-2.2}$\\
FRB 150215 & 782.82 & $0.83\pm0.19$ & $46.5^{+26.6}_{-21.7}$ & $-3.3\pm12.2$ & $3.0\pm27.3$ & $0.1^{+1.3}_{-0.1}$ \\
FRB 150418 & 420.66 & $0.42\pm0.13$ & $60.0^{+33.0}_{-27.6}$ & $36\pm52$ & $-211\pm59$ & $6.1^{+4.7}_{-2.5}$\\
FRB 150807 & 211.01 & $0.17\pm0.08$ & $72.9^{+37.3}_{-32.7}$ & $12.0\pm7$ & $-1.3\pm8$ & $0.03^{+0.16}_{-0.03}$\\
FRB 160102 & 2553.1 & $3.05\pm0.41$ & $21.0^{+12.0}_{-9.8}$ & $-220.6\pm6.4$ &$-249.3\pm12.7$ & $59.5^{+42.2}_{-18.6}$\\
FRB 171209 & 1192.4 & $1.30\pm0.25$ & $37.0^{+21.6}_{-17.4}$ & $121.6 \pm4.2$ & $115.5\pm9.2$ & $4.1^{+3.2}_{-1.8}$\\
FRB 180301 & 240 & $0.20\pm0.10$ & $70.7^{+37.5}_{-32.1}$ & $520-570$ & $504.5-546.2$ & $8.8-11.5$\\
FRB 180309 & 203.42 & $0.17\pm0.08$ & $73.5^{+38.0}_{-33.1}$ & $<150$ & $142.8<$ & $<2.8^{+2.0}$\\
FRB 180311 & 1508.9 & $1.67\pm0.41$ & $31.8^{+21.2}_{-15.6}$ & $4.8\pm7.3$ & $-17.5\pm10.3$ & $1.8^{+1.7}_{	-1.2}$\\
FRB 180714 & 1214.92 & $1.32\pm0.25$ & $36.6^{+21.7}_{-17.2}$ & $-25.9\pm5.9$ & $55.2\pm21.4$ & $0.3^{+1.7}_{-0.3}$\\
FRB 190303.J1353+48 & 170.4 & $0.11\pm0.02$ & $76.3^{+33.5}_{-32.3}$ & $-504.4\pm0.4$ & $498.5\pm12$ & $9.1^{+6.4}_{-2.7}$\\
FRB 190604.J1435+53 & 498.65 & $0.51\pm0.15$ & $56.4^{+32.3}_{-26.3}$ & $-16\pm1$ & $15.7\pm1.4$ & $0.7^{+0.5}_{-0.2}$\\
FRB 190608 &282.08 & $0.25\pm0.10$ & $67.8^{+36.2}_{-30.9}$ & $353\pm2$ & $370.3\pm9.6$ & $8.4^{+6.0}_{-2.6}$\\
FRB 190611 & 247.73 & $0.24\pm0.10$ & $70.2^{+37.8}_{-32.1}$ & $20\pm4$ & $15.9\pm12.1$ & $0.3^{+0.4}_{-0.3}$\\\
FRB 190711 & 520.49 & $0.21\pm0.10$ & $55.6^{+32.0}_{-26.0}$ & $9\pm2$ & $56.05\pm16$ & $1.9^{+1.5}_{-0.8}$\\
FRB 191108 & 515.1 & $0.53\pm0.16$ & $55.7^{+32.1}_{-26.0}$ & $474\pm3$ & $497.7\pm22.6$ & $16.8^{+11.9}_{-5.0}$\\
\hline \hline
\end{tabular}
\label{tab2}
\end{center}
$^{\rm a}${If we assume $\rm DM_{halo}=30\,pc\,cm^{-3}$, the YMW16 model places FRB 180916.J0158+65 within the Milky Way halo.\\}
\end{table*}

\section{Implications for the models of FRBs}

The results presented above can shed light on the origin of FRB 121102. Both FRB 121102 and PSR J1745--2900 have abnormally large 
$\langle B_{\parallel}\rangle$ in their respective categories. The abnormally large RM and $\langle B_{\parallel}\rangle$ of PSR J1745--2900 among the Galactic magnetars is attributed to its proximity to the Galactic Centre. It is therefore tempting to attribute the abnormally large RM and $\langle B_{\parallel}\rangle$ of FRB 121102 to its special environment, likely a putative supermassive black hole in the host galaxy. Indeed, \cite{Zhang18b} suggested that a neutron star whose magnetosphere is sporadically reconfigured by a supermassive black hole can be the source of repeating bursts. This scenario predicts about 50\% of the orbital phase to be active, which seems to be consistent with the recent periodicity search from the source \citep{Cruces20,Rajwade20}.

Another interpretation is to invoke a young magnetar whose MWN provides the strong field to account for the abnormally large RM and $\langle B_{\parallel}\rangle$ \citep{Metzger19}. The peculiar host galaxy of FRB 121102 \citep{Metzger17,Nicholl17,Li19} is consistent with that of a long GRB or superluminous supernova that may give birth to a young magnetar. However, if this is the case, the magnetar that powers FRB 121102 bursts must be much younger or more magnetized than known Galactic magnetars and any other magnetars that power other FRBs. One argument in favor of this is that FRB 121102 seems to be a very active repeating FRB source \citep[e.g.][]{Palaniswamy18,Caleb19}. However, FRB 121102 is not the most active repeater detected by the CHIME collaboration \citep{CHIME121102}. According to this logic, other CHIME repeaters should have greater RM and $\langle B_{\parallel}\rangle$ values, but this is not what is observed.
In fact, the most active FRB source reported by CHIME, FRB 180916.J0158+65 \citep{theCHIME19}, has modest RM and $\langle B_{\parallel}\rangle$ values (see Tables \ref{tab1} and \ref{tab2}). In addition, FRB180301, a recently identified repeater by FAST \citep{Luo20} with a high burst rate, displays a moderate RM.

Assuming that the measured RMs of FRBs do not carry a selection effect (e,g, most FRBs do not yet have polarization measurements), the $\langle B_{\parallel}\rangle$ distribution of Galactic magnetars is not only consistent with that of Galactic pulsars, but also with that of FRBs (Fig. \ref{fig3}).
This may tentatively suggest a magnetar connection of FRBs.
Recently, an FRB-like burst was detected from a Galactic magnetar \citep{STARE2,CHIME20}, suggesting a magnetar-origin of at least some FRBs. We caution that more data are needed (from both the Galactic magnetar side and the FRB side) to make a connection between magnetars and an extreme magneto-ionic environment. For example, the youngest Galactic radio magnetar PSR J1550--5418 has the second strongest surface magnetic field in our sample after PSR J1745--2900. Its $\langle B_{\parallel}\rangle$ is relatively high. However, the entire region within $5\rm\,deg^2$ of the source also has an extraordinarily large absolute value of RM \citep{Oppermann15}, so the extreme magneto-ionic environment is not necessarily provided by its local environment. 

A galactic centre interpretation of the relatively large $\langle B_{\parallel}\rangle$ for {\em all} FRBs is already disfavored by the host galaxy data of a few FRBs that showed that the FRB locations in their host galaxies are usually off center \citep{Bannister19,Ravi19,Marcote20}.

Another interesting observation is that the repeaters in our sample (FRB 121102 excluded) are not systematically more magnetized than apparently non-repeating (one-off) FRBs. This is very likely due to that most non-repeating FRBs are actually repeaters. If most of them are due to a different type of progenitor system, that system should also produce a similar magneto-ionic environments as repeaters. 

\section{Summary}

We have investigated the magneto-ionic environments of the Galactic pulsars/magnetars and FRBs by making use of the measured RM and DM from these sources. 
We investigated the $\langle B_{\parallel}\rangle$ of magnetars as a function of age or surface magnetic field strength and find no apparent trend. 
The $\langle B_{\parallel}\rangle$ of pulsars can be well fitted by a log normal distribution with a mean value of $1.00^{+1.51}_{-0.60}\,\rm\mu G$. The mean $\langle B_{\parallel}\rangle$ of Galactic magnetars (except PSR J1745--2900 at the Galactic Centre) is $1.70\,\rm\mu G$, consistent with the distribution of pulsars.

The $\langle B_{\parallel}\rangle$ distribution of FRB sources is investigated and fitted with a lognormal function.
The mean value of $\langle B_{\parallel}\rangle$ derived by the two methods are $1.77^{+9.01}_{-1.48}\,\rm\mu G$ (NE2001) and $1.74^{+14.82}_{-1.55}\,\rm\mu G$ (YMW16).
The $\langle B_{\parallel}\rangle$ of FRBs is also consistent with that of Galactic magnetars.
FRB 121102 has an extraordinary excess from the FRB $\langle B_{\parallel}\rangle$ distribution. 

The possible origin of the strong Faraday screen of FRB 121102 is discussed in the frameworks of both a supermassive black hole and an MWN. In connection with PSR J1745--2900, the latter possibility is tempting. The magnetar model  requires extreme conditions for the source of FRB 121102. In general, magnetars behind all FRBs (both repeating and appreantly non-repeating) remains a plausible possibility.

\section*{Acknowledgements}

We are grateful to Anda Chen, He Gao, Jinlin Han, Kejia Lee, Rui Luo and Jumei Yao for helpful discussion and comments. W.-Y.W. and X.L.C. acknowledge the support of the NSFC Grants 11633004, 11653003, the CAS grants QYZDJ-SSW-SLH017, and CAS XDB 23040100, and MoST Grant 2018YFE0120800, 2016YFE0100300, R.X.X. acknowledges the support of National Key R\&D Program of China (No. 2017YFA0402602), NSFC 11673002 and U1531243, and the Strategic Priority Research Program of CAS (No. XDB23010200).

\section*{DATA AVAILABILITY}

The data underlying this article are available in the article.

\bibliographystyle{mnras}
\bibliography{reference}

\begin{thebibliography}{}

\bibitem[Arlen et al.(2014)]{Arlen14} Arlen, T.~C., Vassilev, V.~V., Weisgarber, T., et al.\ 2014, \apj, 796, 18

\bibitem[Ando \& Kusenko(2010)]{Ando10} Ando, S., \& Kusenko, A.\ 2010, \apjl, 722, L39

\bibitem[\protect\citeauthoryear{Bannister et al.}{2019}]{Bannister19} Bannister K.~W., et al., 2019, arXiv, arXiv:1906.11476

\bibitem[Bochenek et al.(2020)]{STARE2} Bochenek, C.~D., Ravi, V., Belov, K.~V., et al.\ 2020, arXiv e-prints, arXiv:2005.10828

\bibitem[Caleb et al.(2018)]{Caleb18} Caleb, M., Keane, E. F., van Straten, W., et al. 2018, \mnras, 478, 2046

\bibitem[Caleb et al.(2019)]{Caleb19} Caleb, M., Stappers, B.~W., Rajwade, K., et al.\ 2019, \mnras, 484, 5500

\bibitem[Chatterjee et al.(2017)]{Chatterjee17}Chatterjee, S., Law, C. J., Wharton, R. S., et al.\ 2017, \nat, 541, 58

\bibitem[Cruces et al.(2020)]{Cruces20} Cruces, M., Spitler, L.~G., Scholz, P., et al.\ 2020, arXiv:2008.03461

\bibitem[The CHIME/FRB Collaboration et al.(2020)]{CHIME20} The CHIME/FRB Collaboration, :, Andersen, B.~C., et al.\ 2020, arXiv e-prints, arXiv:2005.10324


\bibitem[Cordes \& Lazio(2002)]{NE2001}Cordes J. M., Lazio T. J. W., 2002, preprint (arXiv:e-print)

\bibitem[Connor et al.(2020)]{Connor20}Connor et al. 2020, arXiv:2002.01399

\bibitem[Dai et al.(2002)]{Dai02} Dai, Z.~G., Zhang, B., Gou, L.~J., et al.\ 2002, \apjl, 580, L7

\bibitem[Day et al.(2020)]{Day20} Day, C.~K., Deller, A.~T., Shannon, R.~M., et al.\ 2020, \mnras, 497, 3335

\bibitem[Deng \& Zhang(2014)]{Deng14} Deng, W., \& Zhang, B.\ 2014, \apjl, 783, L35

\bibitem[Desvignes et al.(2018)]{Desvignes18} Desvignes, G., Eatough, R.~P., Pen, U.~L., et al.\ 2018, \apjl, 852, L12

\bibitem[Dermer et al.(2011)]{Dermer11} Dermer, C.~D., Cavadini, M., Razzaque, S., et al.\ 2011, \apjl, 733, L21

\bibitem[Dolag et al.(2015)]{Dolag15} Dolag K., Gaensler B. M., Beck A. M., Beck M. C., 2015, MNRAS, 451, 4277

\bibitem[Eatough et al.(2013)]{Eat13} Eatough, R.~P., Falcke, H., Karuppusamy, R., et al.\ 2013, \nat, 501, 391 

\bibitem[Fonseca et al.(2020)]{Fonseca20} Fonseca, E., Andersen, B.~C., Bhardwaj, M., et al.\ 2020, arXiv e-prints, arXiv:2001.03595

\bibitem[Gillessen et al.(2017)]{Gillessen17} Gillessen, S., Plewa, P.~M., Eisenhauer, F., et al.\ 2017, \apj, 837, 30

\bibitem[Han \& Qiao(1994)]{Hanqiao94} Han, J.~L., \& Qiao, G.~J.\ 1994, \aap, 288, 759

\bibitem[Han et al.(2006)]{Han06} Han, J.~L., Manchester, R.~N., Lyne, A.~G., et al.\ 2006, \apj, 642, 868

\bibitem[Han et al.(2018)]{Han18} Han, J.~L., Manchester, R.~N., van Straten, W., et al.\ 2018, \apjs, 234, 11

\bibitem[Hessels et al.(2019)]{Hessels19} Hessels, J.~W.~T., Spitler, L.~G., Seymour, A.~D., et al.\ 2019, \apj, 876, L23

\bibitem[Josephy et al.(2019)]{CHIME121102} Josephy, A., Chawla, P., Fonseca, E., et al.\ 2019, \apjl, 882, L18


\bibitem[Kaspi \& Beloborodov(2017)]{Kaspi17} Kaspi, V.~M., \& Beloborodov, A.~M.\ 2017, \araa, 55, 261 

\bibitem[Keane et al.(2012)]{Keane12}Keane, E. F., Stappers, B. W., Kramer, M., \& Lyne, A. G. 2012, \mnras, 425, L71

\bibitem[Keane et al.(2016)]{Keane16} Keane, E.~F., Johnston, S., Bhandari, S., et al.\ 2016, \nat, 530, 453



\bibitem[Li et al.(2019)]{Li19} Li, Y., Zhang, B., Nagamine, K., et al.\ 2019, \apjl, 884, L26

\bibitem[Li et al.(2020a)]{HXMT20} Li, C.~K., Lin, L., Xiong, S.~L., et al.\ 2020a, arXiv e-prints, arXiv:2005.11071

\bibitem[Li et al.(2020b)]{Li20} Li, Z., Gao, H., Wei, J.-J., et al.\ 2020b, \mnras, doi:10.1093/mnrasl/slaa070

\bibitem[{{Lorimer} {et~al.}(2007){Lorimer}, {Bailes}, {McLaughlin}, {Narkevic}, \& {Crawford}}]{Lorimer07}{Lorimer}, D.~R., {Bailes}, M., {McLaughlin}, M.~A., {Narkevic}, D.~J., \&
  {Crawford}, F. 2007, Science, 318, 777
  
\bibitem[Lower et al.(2020)]{Lower20} Lower, M.~E., Shannon, R.~M., Johnston, S., et al.\ 2020, arXiv e-prints, arXiv:2004.11522
  
\bibitem[Luo et al.(2018)]{Luo18} Luo, R., Lee, K., Lorimer, D.~R., et al.\ 2018, \mnras, 481, 2320

\bibitem[Luo et al.(2020)]{Luo20} Luo, R., et al. 2020, Nature accepted

\bibitem[Macquart et al.(2020)]{Macquart20} Macquart, J.-P., Prochaska, J.~X., McQuinn, M., et al.\ 2020, \nat, 581, 391

\bibitem[Manchester(1972)]{Manchester72} Manchester, R.~N.\ 1972, \apj, 172, 43

\bibitem[Manchester(1974)]{Manchester74} Manchester, R.~N.\ 1974, \apj, 188, 637

\bibitem[Manchester et al.(2005)]{Manchester05} Manchester, R.~N., Hobbs, G.~B., Teoh, A., \& Hobbs, M.\ 2005, \aj, 129, 1993 

\bibitem[Marcote et al.(2017)]{Marcote17} Marcote, B., Paragi, Z., Hessels, J.~W.~T., et al.\ 2017, \apjl, 834, L8 

\bibitem[Marcote et al.(2020)]{Marcote20} Marcote, B., Nimmo, K., Hessels, J.~W.~T., et al.\ 2020, \nat, 577, 190

\bibitem[Margalit \& Metzger(2018)]{Margalit18} Margalit, B., \& Metzger, B.~D.\ 2018, \apjl, 868, L4

\bibitem[Margalit et al.(2019)]{Margalit19} Margalit, B., Berger, E., \& Metzger, B.~D.\ 2019, \apj, 886, 110

\bibitem[Masui et al.(2015)]{Masui15} Masui, K., Lin, H.-H., Sievers, J., et al. 2015, Nature, 528, 523

\bibitem[McQuinn(2014)]{McQuinn14} McQuinn, M.\ 2014, \apjl, 780, L33

\bibitem[Mereghetti et al.(2020)]{integral20} Mereghetti, S., Savchenko, V., Ferrigno, C., et al.\ 2020, arXiv e-prints, arXiv:2005.06335 

\bibitem[Metzger et al.(2017)]{Metzger17} Metzger, B.~D., Berger, E., \& Margalit, B.\ 2017, \apj, 841, 14

\bibitem[Metzger et al.(2019)]{Metzger19} Metzger, B.~D., Margalit, B., \& Sironi, L.\ 2019, \mnras, 485, 4091 

\bibitem[Michilli et al.(2018)]{Michilli18} Michilli, D., Seymour, A., Hessels, J.~W.~T., et al.\ 2018, \nat, 553, 182 

\bibitem[Nicholl et al.(2017)]{Nicholl17} Nicholl, M., Guillochon, J., \& Berger, E.\ 2017, \apj, 850, 55


\bibitem[Noutsos et al.(2008)]{Nou08} Noutsos, A., Johnston, S., Kramer, M., et al.\ 2008, \mnras, 386, 1881

\bibitem[\protect\citeauthoryear{Oppermann, et al.}{2015}]{Oppermann15} Oppermann N., et al., 2015, A\&A, 575, A118

\bibitem[Os{\l}owski et al.(2019)]{Osl19} Os{\l}owski, S., Shannon, R.~M., Ravi, V., et al.\ 2019, \mnras, 488, 868


\bibitem[Petroff et al.(2016)]{Petroff16} Petroff, E., Barr, E.~D., Jameson, A., et al.\ 2016, \pasa, 33, e045

\bibitem[Palaniswamy et al.(2018)]{Palaniswamy18} Palaniswamy, D., Li, Y., \& Zhang, B.\ 2018, \apjl, 854, L12

\bibitem[Petroff et al.(2017)]{Petroff17} Petroff, E., Burke-Spolaor, S., Keane, E.~F., et al.\ 2017, \mnras, 469, 4465

\bibitem[Petroff et al.(2019)]{Petroff19} Petroff, E., Hessels, J.~W.~T., \& Lorimer, D.~R.\ 2019, \aapr, 27, 4

\bibitem[Piro \& Gaensler(2018)]{Piro18} Piro, A.~L., \& Gaensler, B.~M.\ 2018, \apj, 861, 150

\bibitem[Planck Collaboration et al.(2018)]{Planck18} Planck Collaboration, Aghanim, N., Akrami, Y., et al.\ 2018, arXiv e-prints, arXiv:1807.06209 

\bibitem[Prochaska et al.(2019)]{Prochaska19} Prochaska, J.~X., Macquart, J.-P., McQuinn, M., et al.\ 2019, Science, 366, 231


\bibitem[\protect\citeauthoryear{Ravi, et al.}{2016}]{Ravi16} Ravi, V., Shannon, R. M., Bailes, M., et al. 2016, Science, 354, 1249

\bibitem[\protect\citeauthoryear{Ravi, et al.}{2019}]{Ravi19} Ravi V., et al., 2019, Natur, 572, 352

\bibitem[Rajwade et al.(2020)]{Rajwade20} Rajwade, K. M., Mickaliger, M. B., Stappers, B. W.\ 2020, arXiv e-prints, arXiv:2003.03596

\bibitem[Ridnaia et al.(2020)]{konus20} Ridnaia, A., Svinkin, D., Frederiks, D., et al.\ 2020, arXiv e-prints, arXiv:2005.11178

\bibitem[Sotomayor-Beltran et al.(2013)]{S-B13} Sotomayor-Beltran, C., Sobey, C., Hessels, J.~W.~T., et al.\ 2013, \aap, 552, A58

\bibitem[Tavani et al.(2020)]{agile20} Tavani, M., Casentini, C., Ursi, A., et al. 2020, arXiv e-prints,
arXiv:2005.12164

\bibitem[Taylor et al.(2009)]{Taylor09} Taylor, A.~R., Stil, J.~M., \& Sunstrum, C.\ 2009, \apj, 702, 1230

\bibitem[Tendulkar et al.(2017)]{Ten17} Tendulkar, S.~P., Bassa, C.~G., Cordes, J.~M., et al.\ 2017, \apj, 834, L7

\bibitem[The CHIME/FRB Collaboration et al.(2019)]{theCHIME19} The CHIME/FRB Collaboration, :, Andersen, B.~C., et al.\ 2019, arXiv e-prints, arXiv:1908.03507

\bibitem[Thornton et al.(2013)]{Thornton13}Thornton, D., Stappers, B., Bailes, M., et al. 2013, Science, 341, 53

\bibitem[Xu \& Han(2015)]{Xu15} Xu, J., \& Han, J.~L.\ 2015, Research in Astronomy and Astrophysics, 15, 1629

\bibitem[Yang \& Zhang(2017)]{Yang17} Yang Y.-P., Zhang B., 2017, \apj, 847, 22


\bibitem[Yao et al.(2017)]{YMW16}Yao, J. M., Manchester, R. N., Wang, N. 2017, \apj, 835, 29

\bibitem[Zhang(2018a)]{Zhang18a} Zhang, B.\ 2018a, \apjl, 854, L21

\bibitem[Zhang(2018b)]{Zhang18b} Zhang, B.\ 2018b, \apjl, 867, L21


\bibitem[Zhong et al.(2020)]{Zhong20} Zhong, S. Q., Dai, Z. G., Zhang, H. M., et al. 2020, arXiv e-prints, arXiv:2005.11109

\bibitem[Zhou et al.(2020)]{Zhou20} Zhou, P., Zhou, X., Chen, Y., et al.\ 2020, arXiv:2005.03517


\end{thebibliography}

\end{document}